\documentclass[aps, prd, twocolumn, lengthcheck, superscriptaddress, showpacs, letterpaper, nofootinbib]{revtex4-1}
\usepackage{amsmath,amssymb}
\usepackage{epsfig}
\usepackage{graphicx}
\usepackage{amsmath}
\usepackage{slashed}
\usepackage{amsfonts}
\usepackage{epstopdf}
\usepackage{color}

\newcommand{\be}{\begin{equation}}
\newcommand{\ee}{\end{equation}}
\newcommand{\bear}{\begin{eqnarray}}
\newcommand{\eear}{\end{eqnarray}}
\newcommand{\ba}{\begin{array}}
\newcommand{\ea}{\end{array}}

\def\be{\begin{eqnarray}}
\def\ee{\end{eqnarray}}

\def\roughly#1{\mathrel{\raise.3ex\hbox{$#1$\kern-.75em%
\lower1ex\hbox{$\sim$}}}}

\def\gsim{\roughly>}

\begin{document}

\title{Baryon as a Quantum Hall Droplet and the Cheshire Cat Principle}

\author{Yong-Liang Ma}
\email{yongliangma@jlu.edu.cn}
\affiliation{Center for Theoretical Physics and College of Physics, Jilin University, Changchun, 130012, China}

\author{Maciej A. Nowak}
\email{nowak@th.if.uj.edu.pl}
\affiliation{M. Smoluchowski Institute of Physics and Mark Kac Complex Systems Research Center, Jagiellonian University, S. \L{}ojasiewicza 11, PL 30-348 Krak\'{o}w, Poland }

\author{Mannque Rho}
\email{mannque.rho@jlu.edu.cn}
\affiliation{Institut de Physique Theorique, CEA Saclay, 91191 Gif-sur-Yvette cedex, France}

\author{Ismail Zahed}
\email{ismail.zahed@stonybrook.edu}
\affiliation{Department of Physics and Astronomy, Stony Brook University, Stony Brook, New York 11794--3800, USA}


\date{\today}
\begin{abstract}
We show that the recent proposal to describe the $N_f=1$  baryon in the large number of color limit
as a quantum Hall droplet, can be understood as a chiral bag in a 1+2 dimensional strip using the
Cheshire cat principle. For a small bag radius, the bag reduces to a vortex line which is the smile
of the cat with flowing gapless quarks all spinning in the same direction. The disc enclosed  by the smile is described by a topological
field theory due to the Callan-Harvey anomaly out-flow. The chiral bag carries naturally unit baryon
number and spin $\frac 12 N_c$. The generalization to arbitrary $N_f$ is discussed.
 \end{abstract}


\maketitle

\setcounter{footnote}{0}


\section{Introduction}
In the large number of color limit, 't Hooft  suggested that QCD is dominated by planar
diagrams, with infinitely many weakly interacting mesons and glueballs~\cite{THOOFT}. Witten argued that in this limit, baryons are heavy solitons made out of the interacting mesons. The coupling of the mesons is weak and of order $1/N_c$, while the coupling of the baryons is strong and of order $N_c$~\cite{WITTEN}.

Chiral solitons made solely of non-linearly interacting pions are prototype of these solitons, an idea put forth
decades ago by Skyrme~\cite{SKYRME} well before the advent of QCD. Chiral solitons are topologically protected in $1+3$
dimensions, and their quantum numbers emerge through semi-classical quantization.  However, their {masses} and
``charges" depend sensitively on the truncated chiral effective action, and somehow less through the more elaborate
chiral holographic  constructions~\cite{HOLO}.

Recently, Komargodski~\cite{ZOHAR} pointed at the peculiar character of the QCD baryons for $N_f=1$ where the chiral effective theory is dominated by the axial U(1) anomaly, and where the soliton construction no longer applies since, for instance,  the topological charge cannot be identified. He noted that the presence of stable superselection rules in the
QCD  vacuum (instanton tunneling between vacua with different Chern-Simons number) implies  the existence of 1+2 dimensional domain walls. These walls connect vacua with different Chern-Simons number and are observed to be stable
at large $N_c$.

Remarkably, when these sheets are finite dimensional with a boundary, Komargodski noted that they can
carry massless edge excitations with baryon quantum numbers. They  are identified with fast spinning baryons. These sheets are  described by
a topological field theory through a level-rank duality argument~\cite{TFT}, much like in the fractional quantum Hall (FQH) effect~\cite{CMHALL}. The
 baryons are analogous to the gapless edge excitations in quantum Hall (QH) droplets. Arguments were put forth for their generalization to arbitrary $N_f$.

In this note, we suggest that these baryonic QH droplets can be understood using the Cheshire cat principle (CCP)~\cite{CHESHIRE}. More specifically, we show that a chiral bag with a single quark species of charge $e$  (electric charge or fermion number) confined to a 1+2 dimensional annulus, leaks most quantum numbers. For all purposes the  bag radius is immaterial thanks to the CCP. In particular, when the bag radius is shrunk to zero, only the smile of the cat is left with spinning gapless  quarks running  luminally, explaining the edge modes and their chirality~\cite{ZOHAR}.

A current transverse to the smile is shown to appear, embodying the Callan-Harvey anomaly out-flow~\cite{CALLAN}. This transverse current is shown to be analogous to the Hall current typical of the QH effect through the emergence of an effective U(1)
gauge field. This U(1)  gauge field lives in the disc enclosed by the Cheshire cat  smile, and is described by a purely topological field theory in 1+2 dimensions. The quantum numbers of this baryon as a QH droplet follow readily from the chiral bag construction. The generalization to many species is discussed.

\begin{figure}[h!]
\begin{center}
 \includegraphics[width=5cm]{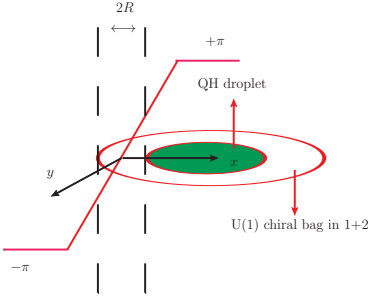}
  \caption{1+2 dimensional chiral bag surrounding a QH droplet. The bag is an annulus of width $2R$ clouded by
   an $\eta^\prime$ with a monodromy of $2\pi$. In the limit of zero bag radius, the chiral bag reduces to a vortex string with unit baryon number.}
    \label{fig_droplet}
 \end{center}
\end{figure}

\section{Bag in a domain  wall}

Consider a 1+2 dimensional chiral bag in the form of an annulus of radius $R$ lying in the $xy$-plane and clouded
by an $\eta^\prime$-field with a monodromy of $2\pi$ or a U(1) winding number of 1.
We will refer to $x$ as the radial direction and to $y$ as the tangential direction as illustrated in Fig.~\ref{fig_droplet}.
The bag consists of free 2-dimensional quarks,  say of charge $e$, and subject to a chiral bag boundary condition along
the radial $x$-direction. We now suggest that this 1+2 dimensional U$(1)$ chiral bag in the limit of zero bag radius is the pancake
baryon suggested by Komargodski thanks to the CCP. Note that in the limit of zero bag radius, the chiral bag
reduces to a vortex line!

\begin{figure}[h!]
\begin{center}
 \includegraphics[width=5cm]{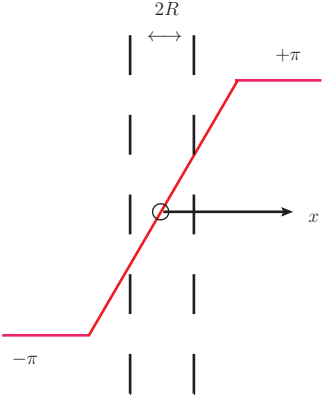}
  \caption{1+2 dimensional chiral bag as an infinite strip in the $y$-direction  (out of the page) with periodic boundary condition
  and $U(1)$ chiral boundary condition along the $x$-direction.}
    \label{fig_wall}
 \end{center}
\end{figure}

The essence of the CCP lies in the fact that the charge $e$ of the chiral bag leaks through an
anomaly. This leakage is best described by noting that in the presence of the $\eta^\prime$-cloud along the
$x$-direction, the Dirac spectrum in the bag undergoes a spectral flow.
Since the discussion is about leakage of charge
along the $x$-direction and flow of charge along the $y$-direction, the shape of the bag as an annulus is topologically
equivalent to an infinite strip along the $y$-direction with periodic boundary condition, and U$(1)$ chiral boundary
condition along the $x$-direction as illustrated in Fig.~\ref{fig_wall}.

For a single quark species, the chiral bag model on the strip is described by
\be
\left(i\partial_t+i\sigma_2\partial_x-i\sigma_3\partial_y\right) q(t,x,y)&=&0\qquad |x|<R\nonumber\\
\left(e^{-i\sigma_2\theta(t,x)}-\sigma_3\epsilon(x)\right)\,q(t, x,y) &=&0\qquad |x|=R,
\label{1}
\ee
with $\epsilon(x)=x/|x|$  the outside normal to the bag, and
 $(\gamma^0,\gamma^1,\gamma^2)=(\sigma_1, i\sigma_3, i\sigma_2)$.
The $\eta^\prime$ field acts only at the boundary through the chiral angle {$\theta=\eta^\prime/f_\eta$}
which is in general time-dependent but $y$-independent. $f_\eta$ is the $\eta^\prime$ decay constant.
Throughout, the reference to chirality in 1+2 dimensions will be a slight abuse of language
for a discrete parity transformation $x_1,x_2\rightarrow -x_1, x_2$, and
$q\rightarrow \sigma_2 q$ with the mass term $\bar q q=q^+\sigma_1 q\rightarrow -\bar q q$  breaking parity.
It only becomes chirality in 1+1 dimensions under dimensional reduction. The anomaly in 1+2 dimensions is the parity anomaly~\cite{RAO}.

With this in mind, the spectral flow is seen by considering
the case of a static boundary condition for the $\eta^\prime$ field. In this case, the mode solution to (\ref{1})
is of the form

\be
q_n(t,x,y)=e^{-iE_nt+ik_yy}\varphi_n(x)
\label{2}
\ee
with $E_n$ following from the transcendental equation

\be
{\rm tan}\left(2R\sqrt{E_n^2-k_y^2}\right)=\frac{1+t_+t_-}{t_-\sqrt{\frac{E_n+k_y}{E_n-k_y}}-t_+\sqrt{\frac{E_n-k_y}{E_n+k_y}}}
\label{3}
\ee
with $t_\pm={\rm tan}(\theta(\pm R)/2))$. Note that the spectrum is now twisted through $t_\pm$.

For the special case of 1+1 dimensions with $k_y=0$, the twist is manifest as (\ref{3}) simplifies to
\be
{\rm tan}(2RE_n)={\rm tan}\left(\frac \pi 2+\frac{\Delta\theta}2\right)
\label{4}
\ee
with $\Delta \theta=(\theta(+R)-\theta(-R))$ as the jump of the $\eta^\prime$-field across the chiral bag.
The twisted spectrum is now

\be
E_n=\frac{(2n+1)\pi}{4R}+\frac {\Delta\theta}{4R}
\label{5}
\ee
with the level $E_{-1}$ crossing zero at the magic angle $\Delta\theta=\pi$ and requiring a vacuum re-definition.
This re-definition implies that the charge in the  chiral bag fractionalizes with the result~\cite{ONE}

\be
\Delta Q=\frac{e\,\Delta \theta}{2\pi}
\label{6}
\ee
as the rest of the charge is now located in the topological charge carried by the outside $\eta^\prime$ field. At the magic angle,
half the charge  is  in  and half is out. The in-charge  is solely carried by the crossing state

\be
q_{-1}=\frac 1{\sqrt{4R}}
\begin{pmatrix}
-i \\
+i
\end{pmatrix}
\label{6X}
\ee
spinning  along the $y$-direction i.e.  $\sigma_2q_{-1}=q_{-1}$  with {\bf fixed helicity}.
 If the monodromy is flipped $2\pi\rightarrow -2\pi$,
the charge and the helicity  are flipped.

The explicit description of the present chiral bag model with space-time dependent boundaries is involved for general $R$
and finite $k_y$,   but around the magic angle  the spectrum becomes CP symmetric  with gapless modes
of energy $E_{-1}\sim k_y$ running in the $y$-direction with fixed helicity.
For this choice, the physics becomes more transparent thanks to the CCP,
with  the emergence of low-dimensional anomalies and bosonization as we now detail.

\section{Anomaly out-flow}

When viewed in 1+1 dimensions, the preceding result is the consequence of an exact bosonization
which captures the essence of the CCP, namely that the bag radius $R$ is immaterial
(the smile of the Cheshire cat). More specifically, (\ref{6}) is the first of the two Abelian
bosonization relations
in 1+1 dimensions

\be
\rho^{1+1}=eq^\dagger q &\rightarrow& \frac{e\partial_x\theta}{2\pi}\equiv \frac{e}{2\pi}\frac{\partial_x\eta^\prime}{f_\eta}\nonumber\\
j^{1+1}_x=eq^\dagger \sigma_3 q &\rightarrow & \frac{e\partial_t\theta}{2\pi}\equiv {\frac{e}{2\pi}\frac{\partial_t\eta^\prime}{f_\eta}}
\label{7}
\ee
These observations are now important for the 1+2 dimensional chiral bag and its mapping on the baryon as a QH droplet.

When the bag radius is increasingly small, the chiral bag is more like a vortex line. At the magic angle, a gapless mode with half fermion number (the other half is sitting on the wall) and momentum $k_y$ flows along the $y$-direction. More importantly, the vortex line carries a charge per unit length $\rho$ and is {\bf leaking radially} a current $j_x$ as given by (\ref{7}) irrespective of how small is $R$! An observer along the vortex line will see $e$-charge increasing or decreasing and would conclude that  his  {\bf tangential} current
$j_y$ is not conserved or anomalous! In other words,
\be
\partial^t\rho +\partial^y j_y=\frac{e^2}{2\pi}E_y
\label{8}
\ee
as if an {\bf emergent}  U$(1)$   {\bf effective electric field} $E_y$ was acting on his vacuum. However, this increase or decrease is caused by the leaking current in the radial direction noted earlier, so that $j_x=j_y$ in magnitude  leading to the identification of the emergent  electric field $E_y$ as
 \be
 \frac {e^2}{2\pi} E_y= \frac{j^{1+1}_x}{L_y}=\frac {e\partial_t\theta}{2\pi L_y}
 \label{9}
 \ee
after using Eqs.~\eqref{7} and \eqref{8} with $L_y$ the $y$-length of the chiral bag as a strip.  A close  reading of (\ref{9}) shows
that $j_x\sim E_y$  which is reminiscent of the Hall current, hence the immediate analogy of the present chiral bag
construction with  the QH effect. This is the Callan-Harvey mechanism for anomaly out-flow~\cite{CALLAN},
now realized for a proposed baryon. It is a physical realisation of the descent  equation between anomalies in even and odd dimensions~\cite{BOOK,BILAL} (and references therein).

\section{Emergent effective action}

The anomaly out-flow to the {\bf outside} disc formed by  the chiral bag as an annulus, can be captured in a 1+2 dimensional
effective action describing the outside of the bag. Indeed, since the leaking and radial current to the chiral bag is $j_x$, its extension in 1+2
dimensions defines the variation of the effective action $S_{1+2}$ with respect to the emergent U(1) gauge field $A_x$ as
\be
\frac{\delta S_{1+2}}{\delta A_x}=\frac{j^{1+1}_x}{L_y}.
\label{10}
\ee
Inserting (\ref{9}) into (\ref{10}) and solving gives
\be
S_{1+2}=\int_{1+2}\frac {e^2}{2\pi}\,A_xE_y=\frac {e^2}{4\pi}\int_{1+2} AdA,
\label{11}
\ee
where covariance was subsumed in the  3-form.
This is  the topological field theory describing the { FQH} droplet outside the bag illustrated in Fig.~\ref{fig_droplet}. One of the chief purpose of the emergent U$(1)$  field $A_\mu$ in the 1+2 dimensional droplet is to enforce the anomaly out-flow, hence its topological rather than dynamical character.

In (\ref{11}) the Chern-Simons coupling or flux attachment factor is $\kappa=e^2/2\pi$. A  coupling of a charge $e$ to the emergent U$(1)$ field in 1+2 dimensions, amounts to a flux attachment of $e/\kappa$. The exchange of any pair of particles will generate a statistical phase $e^2/2\kappa=\pi$ through the Aharonov-Bohm interaction. A charged boson  coupled to the emergent U$(1)$ gauge field in 1+2 dimensions, transmutes to a fermion and vice versa!

The generalization of these results  to many quark species, say of different colors $N_c$, requires the use of non-Abelian
bosonization,  but the CCP still holds~\cite{CHESHIRE,BOOK}. However,  in our case this is not needed. Indeed, ordinary quarks carry baryon or fermion  number $1/{N_c}$ (instead of the  integer $e^2\rightarrow 1$ discussed here), hence a fraction $\pi/N_c$ of the fermion
statistics. This statistics is readily enforced through
a flux attachment factor $\kappa=N_c/2\pi$,  leading to the emergent  Abelian Chern-Simons contribution
\be
\frac {N_c}{4\pi}\int_{1+2} AdA .
\label{12}
\ee
This is the
QH droplet suggested by Komargodski for baryons made of $N_c$ quarks and $N_f=1$. For vanishingly
small radius, the Cheshire cat smile reduces to a vortex line with running gapless quarks all spinning in concert  in the $y$-direction
(recall the magic angle), naturally explaining the
large spin $\frac 12 N_c$. Baryon number is still 1 and now  lodged in the $\eta^\prime$ field through the $2\pi$ monodromy.
Anti-baryons follow from a $-2\pi$ monodromy, with the out-flaw turning to an in-flow.

For arbitrary $N_f$ the spin and statistics arguments do not change as they are solely fixed by $N_c$. However, the leaking flavor currents lead to a U($N_f$) flavor-valued emergent gauge field $\mathbb A_\mu$. Again, the CCP applies {\it mutatis mutandis}. In particular,
the  emergent non-Abelian Chern-Simons action (\ref{12}) is now

\be
\frac {N_c}{4\pi}\int_{1+2} {\rm Tr}\left(\mathbb Ad\mathbb A+\frac 23 \mathbb A^3\right).
\label{13}
\ee

\section{Conclusions}

QCD in the large number of colors and $N_f=1$ does not admit a representation of baryons as chiral solitons since $\pi_3(U(1))=0$. In this limit, Komargodski suggested that baryons are edge excitations of a 1+2 dimensional QH droplet, and concluded that these baryons are heavy and highly spinning.

We have shown that the nature of these baryons follows from an anomaly out-flow (in-flow for anti-baryons)
in  a 1+2 dimensional chiral bag model as an annulus of shrinking size thanks to the CCP. The out-flow from  the bag is captured  by an emergent  U$(1)$
gauge field and described by a topological field theory. The normalization of the latter is  fixed by the quark fractional statistics. The emergence of a QH description in the outside of the bag is an illustration of the Callan-Harvey mechanism for the parity anomaly in 1+2 dimensions.

When the bag is shrunk to zero size, the baryonic charge 1
is lodged in the $2\pi$ monodromy. The chiral bag reduces to a vortex line
(smile of the Cheshire cat), with running gapless modes of fixed helicity as edge excitations carrying net spin $\frac 12 N_c$. These observations generalize to arbitrary $N_f$.

These facts prompt us to ask  about the possible relationship of these highly spinning baryons, with the lowest spinning skyrmions in the spin-isospin tower $J=I=1/2, ..., N_c/2$. As both descriptions rely on QCD in the large number of colors, a dynamical relation may be at work that  selects one from the other. Also, domains of various forms and shapes
made of $\eta^\prime$ or even the lighter $\pi^0$, are likely to form at few times nuclear matter density, say in the crust of neutron stars or deeper, making the baryons as QH droplets potential candidates.

The present interplay between the QH effect and QCD baryons, is much in line with the recent
suggestion between quantum magnetism and QCD confinement~\cite{SULEJMANPASIC}, showing the intricate interplay between concepts of particle physics and condensed matter physics at strong coupling. More insights can be achieved by using perhaps holography, since for instance baryons and the QH states find a common ground for explanation~\cite{HOLO,HALLHOL}.

Finally and more speculatively,  axion quark nuggets are suggested as candidates for dark matter~\cite{ARIEL}. In the cosmic QCD phase transition, axion domain walls are argued to form copiously and decay, trapping anti-matter in the form of 1+3 dimensional nuggets. It is tempting to suggest, that breaking cosmic axion domain walls can also result in 1+2 dimensional pancakes much like the ones discussed here, trapping  topological fields instead, with confined
hypothetical quark fields circling  the boundary. Both the axion (boundary) and the topological fields (disc) are topologically stable and carry energy but are in so far invisible, a good combination for dark matter.
Conversely, 1+3 dimensional $\eta^\prime$ or even neutral $\pi^0$ domain walls instead of axions can be used to trap few quarks in the more standard baryon configuration with low spin, or in the superconducting diquark phase in QCD matter at moderatly high density, with tangible consequences for the neutron star equation of state. We hope to return to these and some other issues next.

\section*{Acknowledgements}

We thank Zohar Komargodski for an inspiring talk.
Three of us (MAN, MR and IZ) would like to thank Jilin University for support where part of this work was started.
 {The work of YM was supported in part by the National Science Foundation of China (NSFC) under Grant No. 11875147 and 11475071.}
The work of MAN  was supported by the Polish National Science Centre (NCN) Grant UMO-2017/27/B/ST2/01139.
The work of IZ was supported by the U.S. Department of Energy under Contract No.
DE-FG-88ER40388.

\section*{Appendix}

In Ref.~\cite{COLOR} a chiral bag model was constructed to prevent the charge  from leaking from the bag following the CCP. In other words a boundary term was added to the chiral bag to seal  the leaking charge. This boundary term can be readily obtained by noting that for $y$-independent fields, (\ref{11}) describes the {\bf outside} of the bag as a line segment in 1+1 dimensions with
\be
 \frac {e}{2\pi}\int_{1+1}Ad\theta=\frac e{2\pi}\int_{1+1} \theta F-\frac e{2\pi}\int_{B}A_0\theta
\label{12X}
\ee
after an integration by parts,
clearly showing the leaking of the $e$-charge through the boundary. To seal the leak, the {\bf inside} of the bag has to be supplemented
by the opposite boundary term

\be
\frac e{2\pi}\int_B nA\frac {\eta^\prime} {f_\eta}  \equiv -\frac e{2\pi}\int_B  \epsilon^{\mu\nu}n_\nu A_\mu\frac {\eta^\prime} {f_\eta}
\label{13X}
\ee
with $n^\nu$ the spatial normal to the bag boundary, after enforcing covariance on the 2-form.
This is exactly the surface term suggested in the Cheshire cat construction in~\cite{BOOK} (see Eq.~(8.24)) and in \cite{CNDIII}. The present arguments illustrate the subtle relationship between the chiral bag in~\cite{COLOR} and the present chiral bag for the baryon as a FQH droplet. In the former the $e$-charge is absolutely confined, while in the latter the $e$-charge is allowed to flow transversely, both making use of a Chern-Simons term. This is the tale of two hotels: the infinite hotel in our world for the confined anomaly, and the finite hotel in the other world for the flowing anomaly. This tale is highly relevant for nuclear and astrophysical processes involving hadron-quark continuity~\cite{CNDIII}. For instance, the role of  the $\eta^\prime$ for the color charge conservation is responsible for the Cheshire Cat mechanism for the {\it tiny} flavor singlet axial charge for the proton $g_A^{(0)}$. Furthermore since the $\eta^\prime$ is expected to become light at high density, it could have a strong impact on the stiffness of the equation of state in compact-star matter required for the observed massive $\gsim 2 M_\odot$ stars.

 \vfil


\begin{thebibliography}{99} \frenchspacing



\bibitem{THOOFT}
  G.~'t Hooft,
  Nucl.\ Phys.\ B {\bf 72} (1974) 461.


\bibitem{WITTEN}
  E.~Witten,
  Nucl.\ Phys.\ B {\bf 160}, 57 (1979).

\bibitem{SKYRME}
  T.~H.~R.~Skyrme,
  Proc.\ Roy.\ Soc.\ Lond.\ A {\bf 260}, 127 (1961);
  I.~Zahed and G.~E.~Brown,
  Phys.\ Rept.\  {\bf 142}, 1 (1986).




\bibitem{HOLO}
  H.~Hata, T.~Sakai, S.~Sugimoto and S.~Yamato,
  Prog.\ Theor.\ Phys.\  {\bf 117}, 1157 (2007)
  [hep-th/0701280 [HEP-TH]];
  D.~K.~Hong, M.~Rho, H.~U.~Yee and P.~Yi,
  Phys.\ Rev.\ D {\bf 76}, 061901 (2007)
  [hep-th/0701276 [HEP-TH]].


\bibitem{ZOHAR}
  Z.~Komargodski,
  arXiv:1812.09253 [hep-th].



\bibitem{TFT}
  P.~S.~Hsin and N.~Seiberg,
  JHEP {\bf 1609}, 095 (2016)
  [arXiv:1607.07457 [hep-th]];
  D.~Gaiotto, Z.~Komargodski and N.~Seiberg,
  JHEP {\bf 1801}, 110 (2018)
  [arXiv:1708.06806 [hep-th]];
  F.~Benini,
  JHEP {\bf 1802}, 068 (2018)
  [arXiv:1712.00020 [hep-th]].



\bibitem{CMHALL}
  D.~Tong,
  arXiv:1606.06687 [hep-th].



\bibitem{CALLAN}
  C.~G.~Callan, Jr. and J.~A.~Harvey,
  Nucl.\ Phys.\ B {\bf 250}, 427 (1985).



\bibitem{CHESHIRE}
  S.~Nadkarni, H.~B.~Nielsen and I.~Zahed,
  Nucl.\ Phys.\ B {\bf 253}, 308 (1985);
  S.~Nadkarni and I.~Zahed,
  Nucl.\ Phys.\ B {\bf 263}, 23 (1986);
  M.~Rho,
  Phys.\ Rept.\  {\bf 240}, 1 (1994)
  [hep-ph/9310300].



\bibitem{RAO}
  A.~J.~Niemi and G.~W.~Semenoff,
  Phys.\ Rev.\ Lett.\  {\bf 51}, 2077 (1983);
  A.~N.~Redlich,
  Phys.\ Rev.\ Lett.\  {\bf 52}, 18 (1984).

\bibitem{ONE}
  I.~Zahed,
  Phys.\ Rev.\ D {\bf 30}, 2221 (1984).




\bibitem{BOOK}
  M.~A.~Nowak, M.~Rho and I.~Zahed,
  {\it Chiral nuclear dynamics} (World Scientific, Singapore,  1996) 528 p

\bibitem{BILAL}
  A.~Bilal,
  arXiv:0802.0634 [hep-th].


\bibitem{COLOR}
  H.~B.~Nielsen, M.~Rho, A.~Wirzba and I.~Zahed,
  Phys.\ Lett.\ B {\bf 269}, 389 (1991);
  H.~B.~Nielsen, M.~Rho, A.~Wirzba and I.~Zahed,
  Phys.\ Lett.\ B {\bf 281}, 345 (1992).

\bibitem{SULEJMANPASIC}
T.~Sulejmanpasic, G.~Shao, A.~W.~Sandvik and M.~Unsal, Phys. Rev. Lett. {\bf 119}, 091601 (2017).


\bibitem{HALLHOL}
 O.~Bergman, N.~Jokela, G.~Lifschytz and M.~Lippert,  JHEP 1010:063 (2010);
  R.~Argurio, M.~Bertolini, F.~Bigazzi, A.~L.~Cotrone and P.~Niro,
  JHEP {\bf 1809}, 090 (2018)
  [arXiv:1806.08292 [hep-th]].









\bibitem{ARIEL}
  S.~Ge, K.~Lawson and A.~Zhitnitsky,
  arXiv:1903.05090 [hep-ph].


 \bibitem{CNDIII} Y.L.  Ma and M. Rho, {\it Effecctive Field Theories for Nuclei and Compact-Star Matter: Chiral Nuclear Dynamics (CND-III)}\
  (World Scientific, Singapore, 2019) Chapter 3.


\end{thebibliography}
\end{document}